# Low-temperature Leidenfrost-like Jumping of Sessile Droplets on Microstructured Surfaces

Wenge Huang [1], Lei Zhao [2, *], Xukun He [1], Yang Li [1], C. Patrick Collier [3], Zheng Zheng [4], Jiansheng Liu [4], Dayrl P. Briggs [3], Jiangtao Cheng [1, *]

[1] Department of Mechanical Engineering, Virginia Tech, Blacksburg, VA 24061, USA

[2] State Key Laboratory of High-performance Precision Manufacturing, Dalian University of Technology, Dalian 116024, PR China

[3] Center for Nanophase Materials Sciences, Oak Ridge National Laboratory, Oak Ridge, TN 37831, USA

[4] School of Electronic and Information Engineering, Beihang University, 37 Xueyuan Road, Beijing 100191, PR China

*Corresponding author: Lei Zhao, leizhao@dlut.edu.cn

Jiangtao Cheng, chengjt@vt.edu

# Abstract

The Leidenfrost effect – the levitation and hovering of liquid droplets on hot solid surfaces – generally requires a sufficiently high substrate temperature to activate liquid vaporization. Here we report the modulation of Leidenfrost-like jumping of sessile water microdroplets on micropillared surfaces at a relatively low temperature. Compared to traditional Leidenfrost effect occurring above 230 °C, the fin-array-like micropillars enables water microdroplets to levitate and jump off the surface within milliseconds at a temperature of 130 °C by triggering the inertia-controlled growth of individual vapor bubbles at the droplet base. We demonstrate that droplet jumping, resulting from the momentum interactions between the expanding vapor bubble and the droplet, can be modulated by tailoring of the thermal boundary layer thickness through pillar height. This enables regulation of the bubble expansion between the inertia-controlled mode and the heat-transfer-limited mode. The two bubble growth modes give rise to distinct droplet jumping behaviors characterized by constant velocity and constant energy regimes, respectively. This heating strategy allows the straightforward purging of wetting liquid droplets on rough or structured surfaces in a controlled manner, with potential applications including the rapid removal of fouling media, even when located in surface cavities.

Research on the Leidenfrost effect dates back to Johann Gottlob Leidenfrost's observation of water droplets' blistering motions on a hot surface in the 18[th] century.[1, 2] Since then, intensive research has addressed this intriguing phenomenon due to its critical importance in various applications such as boiling heat transfer[3], spray cooling[4], electrospray printing,[5, 6] and additive manufacturing[7]. It is widely accepted that the continuous vapor cushion[8] formed beneath the Leidenfrost droplet eliminates the physical contact between the droplet and the surface[9] and consequently minimizes the interfacial hydrodynamic resistance[10] associated with the contact line pinning and solid-liquid friction[11], which is particularly useful for agile droplet manipulations[12, 13] and sustained liquid transport[10, 14]. However, the thermally-insulating vapor cushion also incurs a large thermal resistance[3] and leads to the severe degradation[15] of solid-liquid heat transfer, which is why a substantial surface temperature, *i.e.*, the Leidenfrost point (LFP), must be reached to sustain the intense liquid vaporization essential for droplet levitation. For water droplets on smooth metal surfaces, LFP is usually around 250 °C[10] whereas the effective heat flux is only one third of that on a 110 °C temperature substrate[16, 17]. Therefore, activation of the Leidenfrost-like droplet levitation without sacrificing the heat transfer performance deserves further research. Possible measures include surface engineering and wettability modifications[18] that can alter liquid vaporization and vapor bubble dynamics during the phase-change heat transfer.

As such, achieving maneuverable droplet levitation on hot engineered surfaces will be of great benefit for many applications in highly demanding heat transfer devices[19, 20]. One of the prominent examples is the purging of surface fouling agents[21], *i.e.*, the physical deposition of contaminating particulates on heat exchanging surfaces, which severely impairs the performance of thermal systems like boilers, condensers, and heat exchangers by impeding the effective heat exchange between the working liquids and the solid surface. In particular, this problem becomes further exacerbated for engineered heat transfer surfaces[22] that incorporate corrugations, posts, pyramids, and fins. In this scenario, neither Leidenfrost droplets nor cold droplets at room temperature, such as in spray cooling and tap water rinsing, can effectively remove the sticky fouling materials from surface roughness and cavities. The presence of a continuous vapor layer prevents Leidenfrost droplets from penetrating surface structures and contacting the fouling particles, whilst factors such as contact line pinning and interfacial friction also hinder the capability of cold droplets to effectively dislodge contaminants from wetted surfaces[23]. Methods to address these problems include the redesign of surface textures that allows the controllable wetting[24] and self-purging[23] of sessile droplets in a controllable manner, *i.e.*, the lotus effect, but the problem still remains a challenge.

In this work, we report a new discovery where controllable levitation and purging of sessile Wenzel-state droplets can be achieved at a temperature well below traditional LFP on an engineered surface decorated with micropillar arrays. Here we show that a fin-like micropillar array that penetrates into the droplet base significantly enhances the solid-liquid heat transfer and facilitates the propagation of the thermal boundary layer (TBL)[25], which fosters a superheated environment for inertia-controlled bubble growth[26]. The immense momentum gained from the droplet-bubble interaction eventually stimulates prompt droplet levitation and instantaneous jumping at a markedly low temperature of 130 °C. Decreasing the micropillar height switches the vapor bubble growth to a heat-transfer-limited mode[27] where the otherwise continuous bubble growth is interrupted by vapor condensation at the bubble-droplet interface, causing significant and prolonged vibration of a droplet prior to its jumping. We reveal the mechanisms underlying the tunable Leidenfrost-like droplet jumping by developing appropriate physical models to elucidate the roles of fine surface structures on bubble dynamics. Importantly, we demonstrate its applications in the facile removal of fouling materials from the cavities of hot engineered surfaces.

The surface, which consists of a square lattice of round posts with uniform diameter ($D = 20\,\mu m$), post-to-post spacing ($L = 120\,\mu m$) and height ($H = 80\,\mu m$), hereafter named as $[D,\ L,\ H] = [20, 120, 80]\,\mu m$, was fabricated on a silicon wafer by means of photolithography and deep reactive ion etching.[28] A thin layer of fluoropolymer was then applied via spin-coating to lower the surface energy of the substrate. The droplet levitation experiment was conducted by recording the dynamic behaviors of a sessile water droplet (diameter $D_d \sim 2$ mm) by a high-speed camera at the rate of 10000 frames per second (Extended Data Fig. 1). At room temperature, the droplet remained in a stable Wenzel state[29] with a static contact angle of $118° \pm 1°$ and a sliding angle of $67° \pm 1°$ (Extended Data Fig. 2), due to the sparse distribution of micropillars. After the sessile droplet and the substrate were gently transferred onto a hot plate maintained at 130 °C (Extended Data Fig. 3), the droplet could be levitated and jump off in a Leidenfrost-like manner, *i.e.*, prompt droplet levitation enabled by the liquid vaporization.

The "cold" Leidenfrost phenomenon on modestly heated substrates has been reported in the past[11, 30, 31, 32]. However, what distinguishes our observations from those reported previously is that the low surface superheat utilized here not only triggers the non-wetting transition, but also gives rise to prompt droplet jumping. Fig. 1a shows selected snapshots of the Leidenfrost-like prompt jumping of a sessile water microdroplet. To investigate the kinematic dynamics of the droplet, the onset of the droplet shape deformation, resulting from the capillary perturbations caused by surface heating, was defined as $t = 0$ ms. After that, the droplet was levitated and directly jumped off the substrate within only 1.33 ms. In contrast to the conventional thermally-driven droplet actuations, as in the case of trampolining drops in the traditional Leidenfrost effect[33], the droplet in this scenario exhibited an explosive upward motion. In Fig. 1b, its center of mass monotonically rose with time without any discernible oscillations and eventually reached the maximum jumping height of 2.3 mm at $t = 15.33$ ms, suggesting a distinct mechanism for droplet actuations.

We show that the Leidenfrost-like droplet prompt jumping results from the rectification of the kinetic energy carried by the growing vapor bubble into the upward momentum that is sufficient to lift the entire droplet. At surface temperature $T_w = 130\ °C$, contact boiling occurred at the droplet base[17], leading to the successive nucleation and growth of individual vapor bubbles[27]. Fig. 2a shows that an individual vapor bubble firstly nucleated at $t = 0$ ms. After that, the vapor bubble expanded rapidly and reached the droplet's periphery at $t = 1.32$ ms. The striking coincidence between this timescale and the droplet's dwelling time (1.33 ms) implies that the Leidenfrost-like droplet jumping is contingent on the momentum interactions between this individual vapor bubble and the sessile droplet.

A detailed inspection on the vapor bubble growth in Fig. 2b reveals a bubble expanding velocity as fast as $U_e \approx 4$ m/s, leading to an average Reynolds number of $Re = \frac{U_e D_d}{\nu} \approx 2 \times 10^4$ and Mach number of $Ma = \frac{U_e}{c} \approx 2.6 \times 10^{-3}$, where $\nu$ is water kinematic viscosity and $c$ is the speed of sound in water. Therefore, the effect of viscous dissipation and water compressibility could be neglected, indicating that the vapor bubble interface expansion follows an inertia-controlled mode[27]. This allows us to treat the vapor bubble expansion using the Rayleigh-Plesset equation in a potential flow approach[27]:

$$\rho_l R \ddot{R} + \rho_l \frac{3\dot{R}^2}{2} = \left(P_v - P_\infty - \frac{2\sigma}{R}\right)\sin^2\beta \qquad (1)$$

where $R$ is the bubble contact radius, $\rho_l$ is water density, $P_v$ is pressure inside the bubble, $P_\infty$ is the pressure of bulk water, $\beta$ is the bubble contact angle and $\sigma$ is the surface tension, as shown in Fig. 2c. Eq. (1) describes how the overpressure-induced potential energy stored in the expanding vapor bubble is converted into the kinetic energy of the droplet (Supplementary Discussion 2). The vaporization-induced bubble overpressure could be estimated by the Clausius-Clapeyron equation[27] as $P_v - P_\infty = \left(\frac{\Delta T}{T_{sat}}\right) h_{lv}\rho_v$, where $\Delta T$ is the vapor bubble superheat temperature change, $T_{sat}$ is the water saturation temperature, $h_{lv}$ is specific evaporation enthalpy, and $\rho_v$ is the vapor density. In general, $\Delta T$ would vary with time and the position in the bubble, but a constant $\Delta T$ could be assumed in the case of a strong thermal diffusion that makes the temperature gradient inside the bubble negligible[34]. This approximation holds true as long as the characteristic diffusion length $L_H \sim \sqrt{2\alpha_v t}$ is larger than the bubble size $R$, where $\alpha_v$ is the thermal diffusivity of water vapor. Using $\Delta T = 30$ K gives rise to a rough estimation of $P_v - P_\infty \approx 1.08 \times 10^5$ Pa. As shown in Fig. 2b, the detected vapor bubble growth started from $R = 90$ μm, yielding the Laplace pressure $\frac{2\sigma}{R} \leq 1.31 \times 10^3$ Pa. Therefore, the Laplace pressure term in Eq. (1) is neglected. In the first stage of the bubble expansion with a constant $\Delta T$, Eq. (1) can be solved analytically as:

$$R\left(t\right) = \left[\frac{2}{3}\left(\frac{\Delta T}{T_{sat}(P_a)}\right)\frac{h_{lv}\rho_v}{\rho_l}\right]^{\frac{1}{2}} \sin\beta \cdot t \tag{2}$$

Such a linear increase of $R(t)$ with time implies that the bubble expansion would eventually surpass the thermal diffusion[34], which has the time dependence $R \sim t^{0.5}$, after which the cooling effect of vaporization[34] causes $\Delta T$ to rapidly deteriorate at the liquid-vapor interface. The overpressure would quickly relax to zero[35], resulting in $P_v - P_\infty - \frac{2\sigma}{R} \approx 0$. This prediction aligns well with the sharp turning of the bubble radius curve at $t_1 \approx 0.1$ ms in Fig. 2b. At this moment, the bubble expansion was solely sustained by the droplet inertia and the total kinetic energy of the bubble-droplet system was conserved: $\frac{d}{dt}\left(\rho_l R^3 \dot{R}^2\right) = 0$. The contact bubble radius $R$ for $t > t_1$ can be solved as:

$$R = \left(R_1^{3/2}\dot{R}_1\, t\right)^{0.4} \tag{3}$$

where $R_1$ is the contact radius at $t_1$. Fig. 2b shows a remarkable agreement between the experimental results of bubble expansion and the two-stage theoretical model proposed by us, further confirming the dominant role of the overpressure and inertia effect in controlling the bubble growth for Leidenfrost-like droplet jumping.

The inertia-controlled bubble growth contributes to the droplet levitation by its momentum exchange with the water droplet. The propulsive force is obtained by taking the derivative of the droplet's upward momentum $M_z$ (Supplementary Discussion 3).

$$F_z = \frac{dM_z}{dt} \approx \frac{\pi \rho_1 R^2 \dot{R}^2 (4 - 3\cos\beta)\cos\beta}{\sin^4\beta} \qquad (4)$$

Combining Eq. (3) and Eq. (4) gives an estimated value for $F_z \approx 2.44 \times 10^{-3}$ N. As illustrated in Fig. 2d, the droplet gravity is $G = 4.16 \times 10^{-5}$ N, which is orders of magnitude smaller than $F_z$. Notably, the growing vapor bubble gradually separates the physical contact between the droplet base and the substrate, causing a continually decreasing surface adhesion. Therefore, with the rectified kinetic energy overwhelming the resistance, the water droplet became levitated and instantaneously jumped off the substrate without apparent oscillations, in a Leidenfrost-like manner but at a markedly low temperature of 130 °C.

We find the Leidenfrost-like droplet jumping breaks down on substrates with shorter pillars as the droplet-vapor dynamics become different. Fig. 3a presents the selected snapshots of a droplet's actuation on a substrate with short micropillars ( $H = 20$ μm). In contrast to the direct Leidenfrost-like jumping, the droplet experienced substantial vertical stretching and vibrations until its ultimate jump off with an extended dwelling time of 941 ms, which is >700 times greater than that for Leidenfrost-like jumping. In Fig. 3b, the droplet actuations could be generally divided into two stages. Before $t = 600$ ms, the droplet fluctuated randomly with a small magnitude and no apparent periodicity. After $t = 600$ ms, the oscillations became more pronounced with an evident frequency of 41.6 Hz, which is consistent with the characteristic frequency of a water spring in bouncing drops[36]. Therefore, we term this phenomenon as vibrational droplet jumping, as the droplet's dwelling time is increased by hundreds of times due to the prolonged vibrations.

To unveil the mechanism of vibrational droplet jumping, we investigated the dynamics of individual growing bubble as represented by the snapshots in Fig. 4a. Initially, the vapor bubble followed an inertia-controlled expansion (Fig. 4b) because the bubble expansion is always initiated by the vaporization-induced overpressure. However, the bubble growth in vibrational droplet jumping was interrupted by an apparent bubble shrinking process (from 0.9 ms to 4 ms), causing the bubble radius to decrease for several milliseconds and then rise again as shown in Fig. 4b (Extended Data Fig. 4). This intermittent shrinking is attributed to the limited propagation of the TBL, within which the liquid becomes superheated due to surface heating but remains subcooled elsewhere. For the thermal interaction between the sessile droplet and the heating substrate, the thermal timescale[37] can be estimated as $\tau_{th} = \frac{K_w \rho_1 c_p}{h_c^2} \approx 25$ ms with $K_w$ being the thermal conductivity of liquid water, $c_p$ being the thermal capacity of liquid water and $h_c \approx 10000$ W/(m$^2$·K) being the estimated convective heat transfer coefficient between the droplet and the substrate[27]. The Fourier number characterizing the extent of heat conduction is estimated to be $Fo = \frac{\alpha_w \tau_{th}}{D_d^2} = 1.05 \times 10^{-3}$, verifying the presence of a thin TBL with a steep temperature gradient. Here $\alpha_w$ is the thermal diffusivity of liquid water. Then the TBL propagation velocity $v_{TBL}$ can be further evaluated by solving the heat transfer equation at the droplet base (Supplementary Discussion 4).

$$v_{TBL} = \text{erf}^{-1}\left(\frac{T_{sat}-T_w}{T_0-T_w}\right)\sqrt{\frac{\alpha_w}{t}} \tag{5}$$

where $T_0 = 20\,°C$ is the initial temperature of water. Specifically, the characteristic TBL propagation velocity can be evaluated as $v_{c,TBL} \approx 6.4 \times 10^{-4}$ m/s by substituting $t = \tau_{th}$. This value is three orders of magnitude smaller than the inertial-controlled bubble expansion $U_{i,e} \approx 4$ m/s. Therefore, the inertial bubble expansion would inevitably surpass the TBL propagation whilst the TBL thickness remains almost constant. As depicted in Fig. 4c, the vapor inside the bubble would condensate once its outer edge breaches the TBL. The bubble growth can continue only if the heat transfer from the substrate eventually surpasses the energy loss via condensation, indicating a heat-transfer-limited mode. Therefore, we applied the energy balance to describe the bubble expansion herein:

$$h_{lv}\rho_v \frac{d}{dt}\left(\frac{\pi}{3}\frac{(2+\cos\beta)(1-\cos\beta)^2}{\sin^3\beta}R^3\right) = 2\pi R^2 q_b \tag{6}$$

where $q_b$ represents the heat flux from the silicon substrate to the bubble base. Specifically, the convective heat flux can be estimated as $q_b = h_c(T_w - T_{sat})$. As a result, the temporal evolution of vapor bubble radius in this stage follows:

$$R(t) \sim \frac{q_b}{h_{lv}\rho_v}\frac{2\sin^3\beta}{(2+\cos\beta)(1-\cos\beta)^2} \cdot t \tag{7}$$

Such a linear growth of bubble with time is validated by our experimental results as shown in Fig. 4b.

The propulsive force provided by the heat-transfer-limited bubble growth can be evaluated by combining Eqs. (4) and (7), yielding $F_z \approx 2.16 \times 10^{-6}$ N. Given the droplet gravity $G = 4.16 \times 10^{-5}$ N, the water droplet could not be completely levitated by an individual expanding bubble. The droplet hence entered the trampolining mode with strong capillary oscillations caused by droplet-bubble interactions, until it could build up sufficient kinetic energy for subsequent jumping[33, 38, 39].

A detailed analysis on the droplet jumping velocity $v_j$ versus the droplet volume $V_0$ reveals two distinct jumping modes as delineated in Fig. 4d. For Leidenfrost-like droplet jumping, the kinetic energy $E_k$ of the jumping droplet originates from the overpressure potential energy of an individual bubble, which is mainly determined by the superheat[27]. The initial potential energy stored in each individual vapor bubble formed inside droplets with varying volumes can be taken at the same level, considering the droplets are deposited on the substrate with identical surface temperature. As a result, the kinetic energy $E_k \sim \frac{1}{2} \rho_l V_0 v_j^2$ is approximately constant for different droplet volume $V_0$, yielding a scaling of $v_j \sim V_0^{-0.5}$ as manifested by Leidenfrost-like droplet jumping (Fig. 4d). For vibrational jumping on the substrate $[D, L, H] = [20, 120, 20] \, \mu m$, the droplet needs to overcome the gravity $G$ and surface adhesion from the substrate $F_A$. The surface adhesion $F_A$ is proportional to the pillar perimeter[40] $\sigma \pi D$ and the number of pillars under the droplet base $\left(\frac{D_d}{L}\right)^2$, giving $F_A \sim \frac{\sigma \pi D}{L^2} D_d^{\ 2} \approx 1.27 \times 10^{-3}$ N. The droplet gravity $G = 4.16 \times 10^{-5}$ N is thus neglected in the force analysis. The propulsive force is scaled by considering the temporal variation of momentum as $F_z \sim \rho_l D_d^3 \frac{v_j}{D_d/v_j}$. A threshold velocity acquired by $F_z = F_A$ marks the onset of droplet levitation, yielding $v_j \sim \sqrt{\frac{\sigma D}{L^2 \rho_l}}$ for vibrational jumping. This prediction indicates that all water microdroplets jump with a constant velocity, aligning well with the observed jumping velocities as shown in Fig. 4d.

To comprehensively understand the influence of micropillars on the droplet-substrate heat transfer and two distinct jumping modes, we conducted experiments of droplet jumping on substrates with varying temperature $T_w$ and pillar height $H$. The phase map in Fig. 5a depicts three different regimes of droplet behaviors. Increasing $T_w$ above 130 °C gave rise to the vibrational droplet jumping (regime I) and Leidenfrost-like droplet jumping (regime II). We emphasize that the jumping mode is predominantly determined by the micropillar height $H$ and increasing $T_w$ does not necessarily initiate the vibration-to-Leidenfrost-like transition. Specifically, the vibrational droplet jumping occurred for $H < 60 \, \mu m$ even with $T_w = 170$ °C (Extended Data Fig. 5), and the Leidenfrost-like droplet becomes probable only when $H \geq 60 \, \mu m$. This is a strong indication of the fact that increasing $T_w$ does not impact the TBL propagation whilst the structure of the TBL could be significantly altered by the microstructures. Such effect can be examined by revisiting Eq. (5) to estimate the TBL propagation velocity $v_{TBL}$, as shown in Fig. 5b. Increasing $T_w$ from 130 ℃ to 170 ℃ only increases $v_{TBL}$ by 78.6% from $6.4 \times 10^{-4}$ m/s to $1.1 \times 10^{-3}$ m/s, still orders of magnitude smaller than the inertia-controlled bubble expansion ($U_{i,e} \approx 4$ m/s). Therefore, during the short-time inertial bubble expansion that determines the ultimate droplet jumping mode, the TBL remains almost unchanged even when $T_w$ is increased significantly. We conducted heat transfer simulations on COMSOL® 5.6 to investigate the effect of micropillars on the distribution of TBL (Supplementary Discussion 5). Specifically, Fig. 5c shows that the temperature distribution in the vicinity of the substrate. The isothermal contour of $T = 100\ ℃$, which marks the boundary of the TBL, conforms to the profile of micropillars, suggesting that the microstructures act as micro-fins to extend the superheated liquid domain. We define the TBL thickness as $h_{TBL} = \frac{V_{sup}}{L_p}$, where $V_{sup}$ corresponds to the total volume of superheated liquid and $L_p$ is the projected length of the computational domain (Supplementary Discussion 5). We compare $h_{TBL}$ at different $T_w$ and $H$ in Fig. 5d. Increasing $T_w$ from 130 ℃ to 170 ℃ only marginally increases $h_{TBL}$ from 90 μm to 115 μm, aligning well the theoretical prediction of Eq. (5) that $\Delta h_{TBL} = \int_0^{\tau_{th}} [v_{TBL}(T_w = 170\ ℃) - v_{TBL}(T_w = 130\ ℃)] dt \approx 20\ \mu m$. Alternatively, for $T_w = 130\ ℃$, $h_{TBL}$ is substantially increased from 40 μm to 90 μm when the pillar height is increased from 20 μm to 80 μm, confirming that the effect of microstructures in extending the TBL. Moreover, in regime III of Fig. 5a where $T_w$ is below 120 ℃, no vapor bubble can be observed and the droplet would steadily remain in the Wenzel state till complete evaporation, due to the insufficient thermal energy input for bubble formation and growth.

We demonstrate that engineering surface microstructures to manipulate the growth of bubble expansion and droplet jumping behaviours can be employed as an effective strategy for rapid droplet shedding on hot substrates. Figs. 6a and 6b show time-lapsed images of ejection of sessile water droplets on tilted substrates with different pillar heights. For the tilted substrate $[D,\ L,\ H] = [20, 120, 20]$ μm, the vibrating droplet initiated the out-of-plane jumping and then the water droplet landed softly on the substrate remaining in the low-friction Cassie state until it slid off the substrate. For the tilted substrate $[D,\ L,\ H] = [20, 120, 80]$ μm, the explosive droplet jumping caused the droplet to jump off the substrate with a maximum height of 6 mm, which is three times as large as the droplet diameter. Then the droplet experienced repetitive rebounding and falling for several cycles before it finally rolled off the substrate.

The rapid shedding of initially sticky droplets on heated substrates is of particular relevance to the fouling removal on the highly-demanding heat exchanger surfaces. Under spray cleaning or rinsing conditions, neither Leidenfrost nor cold droplets effectively eliminate deposited particulates from surface roughness. A continuous vapor layer restricts Leidenfrost droplets from accessing surface structures, while factors like contact line pinning and interfacial friction impede the ability of cold droplets to dislodge contaminants. Fig. 6c illustrates an alternative process of removing fouling in the interstitial cavities of surfaces by leveraging the explosive droplet jumping discovered by this work. The contaminant used in the experiments was hydrophilic prism polishing powder (~ 3 μm in diameter). To mimic the gradual deposition of particles on the surface of heat exchangers, the powder particles are suspended in a water droplet with later evaporation process to deposit them on the surface, which is analogous to coffee ring effect[41](Supplementary Discussion 6). At first, the droplet was in Wenzel state, which allows it to penetrate into the interstitial cavities to catch the contaminants. As the surface temperature was heated up to a moderate level ($T_w = 130\ °C$), the generation of vapor bubbles effectively dislodged the residual contaminant particles and drove them to suspend in the droplet. Along with the droplet jumping, the fouling even in surface roughness and cavities can be effectively purged in a deep cleaning manner.

We envisage that the above-mentioned strategy for facile actuations of sessile liquid droplets in an ultrafast, yet controlled manner has a wide range of applications in highly demanding heat transfer and fluid manipulation scenarios. Particularly, this study paves a new path for the deep cleaning of fouling settled in surface cavities, a critical factor that tends to lead to the severe performance degradation of engineered surfaces. From a broader perspective, the exploration of minimizing the thermal cost associated with the vapor-mediated droplet actuations represents an important advance in our understanding of the complex transport of momentum, mass and heat in the phase change heat transfer process, enabling the rational design of surfaces with exceptional heat transfer performance, extended durability, and excellent anti-fouling properties.

**Acknowledgements** This work was supported by NSF CBET under grant number 2133017 and NSF ECCS under grant number 1808931. L.Z. acknowledges financial support from the National

Natural Science Foundation of China under grant number 52105174, the Opening Project of the Key Laboratory of Bionic Engineering (Ministry of Education), Jilin University under grant number KF2023004. Device fabrication, and a portion of the analysis and manuscript preparation were performed at the Center for Nanophase Materials Sciences of the Oak Ridge National Laboratory, which is a US DOE Office of Science User Facility.

**Author contributions** J.C. and W.H. conceived the research. J.C. and L.Z. supervised the research. W.H. designed and carried out the experiments, W.H., L.Z. and J.C. analyzed the data and wrote the original manuscript. W.H., X.H., C.P.C., D.P.B., Y.L., Z.Z., J.L. and J.C. prepared the samples. All authors wrote and edited the manuscript.

**Competing interests** The authors declare no competing interests.

**Figure Captions**

**Figure 1. Leidenfrost-like droplet jumping dynamics on hot micropillared surface. a** Selected snapshots of Leidenfrost-like droplet jumping on a micropillared substrate ($[D, L, H] = [20, 120, 80]$ μm) with surface temperature $T_w = 130$ °C. The inset in (**a**) is the scanning electron micrography (SEM) of the micropillared substrate. **b** Height variation of the center of mass of the droplet shown in (**a**). The time $t = 0$ ms denotes the onset of the interfacial deformation. Supplementary Movie S1 provides additional details.

**Figure 2. Rapid vapor bubble expansion for Leidenfrost-like droplet jumping. a** Top-view snapshots of vapor bubble growth on substrate $[D, L, H] = [20, 120, 80]$ μm at 130 °C. The scale bar is 1 mm. Supplementary Movie S2 provides additional details. **b** Temporal evolution of the vapor bubble radius from (**a**). The error bars represent the potential resolution errors. **c** Diagram of vapor bubble expansion via momentum interaction with the surrounding liquid. **d** Equivalent upward force generated by the rapid bubble expansion.

**Figure 3. Droplet vibration jumping dynamics on hot micropillared surface. a** Selected snapshots of droplet's vibrational jumping on the micropillared substrate $[D, L, H] = [20, 120, 20]$ μm at 130 °C. The inset in (**a**) is the SEM image of the micropillared substrate. **b** Height variation of the center of mass of the droplet shown in (**a**). Supplementary Movie S3 provides additional details.

**Figure 4. Vapor bubble shrinking during the vibrational droplet jumping. a** Top-view snapshots of vapor bubble growth on substrate $[D, L, H] = [20, 120, 20]$ μm at 130 °C. **b** Temporal evolution of vapor bubble radius on substrates with different micropillar heights ($H = 20, 60, 80$ μm). The error bars represent the potential resolution errors. Supplementary Movies S4

and S5 provide additional details. **c** Schematic illustrations of superheated interfacial water layer impacting the bubble growth. The growing vapor condenses after meeting the subcooled water outside the TBL. **d** Jumping velocity of droplets with different volumes during vibrational jumping (on substrate $[D,\ L,\ H] = [20, 120, 20]\ \mu m$) and Leidenfrost-like jumping (on substrate $[D,\ L,\ H] = [20, 120, 80]\ \mu m$).

**Figure 5. Effect of micropillar height $H$ and substrate temperature $T_w$ on droplet jumping behaviors. a** Phase map of droplet jumping behaviors on substrates with different micropillar heights and different surface temperatures. **b** Theoretical prediction of the TBL propagation velocity at different substrate temperatures. **c** Simulated results of temperature distribution showing the TBL on substrates with micropillar height from 20 μm to 80 μm. The isothermal contour of 100 °C is donated with white dashed line. **d** Simulated TBL thickness on substrates with different micropillar heights (from 20 μm to 80 μm) and different substrate temperatures (from 130 °C to 170 °C).

**Figure 6. Rapid droplet purging on different substrates and surface deep fouling removal. a** Droplet sliding on substrate $[D,\ L,\ H] = [20, 120, 20]\ \mu m$ at 130 °C. **b** Droplet sliding on substrate $[D,\ L,\ H] = [20, 120, 80]\ \mu m$ at 130 °C. **c** Schematic (top) and experimental snapshots (bottom) of dislodging and removal of fouling from surface roughness by sliding droplet on tilted substrate $[D,\ L,\ H] = [20, 120, 20]\ \mu m$ at 130 °C. All the substrates are tilted at 16°. Supplementary Movies S6 and S7 provide additional details.

**Methods**

**Substrate preparation**

Polished P-type silicon wafers of 100 mm diameter and $550 \pm 25$ μm thickness were used as the substrates in this work. Standard photolithography process was performed with a SUSS MicroTech contact aligner. Then the substrates were etched with Oxford RIE to fabricate the well-defined micropillar arrays. The micro-pillared substrates were conformally coated with fluoropolymer (PFC 1601V, Cytonix Corporation) using a spin coater at 3000 rpm for 30 s and then baked at 100 °C for 1 hour. Substrate micropillar diameter, height and periodicity (pitch-to-pitch distance) are denoted with $D$, $H$ and $L$, respectively. More detailed information about the substrates is given in Extended Data Fig. 2. The surface roughness of the substrate, assessed using atomic force microscopy (AFM, Asylum Jupiter XR), is approximately 7 nm, as illustrated in Supplementary Fig. 2.

**Data availability**

Data analyzed during this study are included in this Article. Source Data are provided with this paper.

**Code availability**

Codes used to generate the data presented in this study are available from the corresponding authors upon reasonable request.